\documentclass[sn-basic]{sn-jnl}
\usepackage{graphicx}%
\usepackage{multirow}%
\usepackage{amsmath,amssymb,amsfonts}%
\usepackage{amsthm}%
\usepackage{mathrsfs}%
\usepackage[title]{appendix}%
\usepackage{xcolor}%
\usepackage{textcomp}%
\usepackage{manyfoot}%
\usepackage{array}          
\usepackage{booktabs}   
\usepackage{longtable}    
\usepackage{etoolbox}
\usepackage{graphicx}     
\usepackage{paralist}      
\usepackage{ragged2e}    
\usepackage{booktabs,makecell, multirow, tabularx}
\usepackage{makecell}   
\usepackage{booktabs}%
\usepackage{algorithm}%
\usepackage{algorithmicx}%
\usepackage{algpseudocode}%
\usepackage{listings}%

\theoremstyle{thmstyleone}%
\theoremstyle{thmstyletwo}%

\theoremstyle{thmstylethree}%

\begin{document}
	
\title[QCIC modeling to optimize therapy for advanced renal cell carcinoma]{Quantitative cancer-immunity cycle modeling to optimize bevacizumab and atezolizumab combination therapy for advanced renal cell carcinoma}
	
	\author[1]{\fnm{Lei} \sur{Du}}\email{dulei@tiangong.edu.cn}
	
	\author[1]{\fnm{Chenghang} \sur{Li}}\email{lichenghang@tiangong.edu.cn}

	\author*[1,2]{\fnm{Jinzhi} \sur{Lei}}\email{jzlei@tiangong.edu.cn}
	
	\affil[1]{\orgdiv{School of Mathematical Sciences}, \orgname{Tiangong University}, \orgaddress{ \city{Tianjin}, \postcode{300387}, \country{China}}}
	
	\affil[2]{\orgdiv{Center for Applied Mathematics}, \orgname{Tiangong University}, \orgaddress{ \city{Tianjin}, \postcode{300387}, \country{China}}}
	
	
\abstract{The incidence of advanced renal cell carcinoma(RCC) has been rising, presenting significant challenges due to the limited efficacy and severe side effects of traditional radiotherapy and chemotherapy. While combination immunotherapies show promise, optimizing treatment strategies remains difficult due to individual heterogeneity. To address this, we developed a Quantitative Cancer-Immunity Cycle (QCIC) model that integrates ordinary differential equations with stochastic modelling to quantitatively characterize and predict tumor evolution in patients with advanced RCC. By systematically integrating quantitative systems pharmacology principles with biological mechanistic knowledge, we constructed a virtual patient cohort and calibrated the model parameters using clinical immunohistochemistry data to ensure biological validity. To enhance predictive performance, we coupled the model with pharmacokinetic equations and defined the Tumor Response Index (TRI) as a quantitative metric of efficacy. Systematic analysis of the QCIC model allowed us to determine an optimal treatment regimen for the combination of bevacizumab and atezolizumab and identify tumor biomarkers with clinical predictive value. This study provides a theoretical framework and methodological support for precision medicine in the treatment of advanced RCC.}
	
	
\maketitle
	
	\newpage
	
\section{Introduction}
\label{sec1}
	
Renal cell carcinoma (RCC) is a malignant tumor originating from the renal tubular epithelium, accounting for approximately 80--90\% of primary renal malignancies. Epidemiological data indicate that the incidence of advanced RCC continues to rise. While traditional standard-of-care modalities such as radiotherapy and chemotherapy can reduce tumor burden in patients with unresectable RCC, they are often limited by low response rates, high recurrence, and significant toxicities. Recently, breakthrough advances in immunotherapy--specifically immune checkpoint inhibitors (ICIs)--have demonstrated encouraging clinical benefits for advanced RCC. Notably, combination strategies involving ICIs and targeted therapies have significantly prolonged overall survival and are now included in standard first-line treatment protocols.
	
Mechanistically, tumor cells facilitate immune escape by upregulating immune checkpoint molecules to suppress T-cell antitumor activity. ICIs can effectively restore T-cell cytotoxicity by blocking these immunosuppressive pathways. Clinical trials have confirmed that PD-L1 inhibitors, such as atezolizumab, provide significant benefits in advanced RCC. Concurrently, vascular endothelial growth factor (VEGF) within the tumor microenvironment promotes angiogenesis by binding to VEGF receptor (VEGFR) on endothelial cells, securing the nutrient supply necessary for tumor growth. Anti-angiogenic agents, such as bevacizumab, inhibit this process by neutralizing VEGF. The clinical value of this combined approach was validated in the IMmotion151 study, in which atezolizumab combined with bevacizumab significantly improved progression-free survival in patients with advanced RCC.
	
Mathematical oncology has recently seen breakthroughs in both theoretical frameworks and practical applications, offering substantial value for clinical translation. Lai et al. utilised reaction-diffusion equations to simulate dynamic interactions among immune cells, cytokines, and tumors \citep{Lai.BMCSystemsBiology.2017,Lai.PNAS.2018,Lai.SciChinaMath.2020} . Their work not only elucidated key immunoregulatory mechanisms but also provided a quantitative basis for optimizing treatment strategies. Similarly, Li et al. systematically characterized tumor-immune interactions using ordinary differential equations (ODEs), offering theoretical support for immunotherapy planning \citep{Li.BullMathBiol.2024}. Zhou et al. developed a multi-compartment model to simulate dynamic immune response during viral infection, offering insights applicable to vaccine development \citep{Li.PLoSCompuBiol.2023}. Furthermore, Wang et al. advanced a quantitative systems pharmacology (QSP) framework, achieving accurate characterization of the spatiotemporal dynamics of immune cells and cytokines via compartmental modeling \citep{Wang.JITC.2021,Wang.iScience.2022,Wang.NPJPrecisOncol.2023}. These quantitative models have successfully aided in the design and optimization of innovative drugs.
	
Building upon previous work \citep{Li.NPJSystBiolAppl.2025}, we further optimized the Quantitative Cancer-Immunity Cycle (QCIC) model to simulate treatment responses in RCC patients and predict clinical benefits. This model integrates systems biology, immunology, and quantitative pharmacology into a multiscale framework that systematically depicts the tumor-immunity cycle. The optimized QCIC model comprises five compartments: (1) immune cell generation (bone marrow and thymus), (2) circulatory system (peripheral blood), (3) immune activation (tumor-draining lymph nodes), (4) effector compartment (tumor microenvironment), and (5) antigen transport (lymphatic vessels). We detailed the dynamic interactions among immune cell populations--including na\"{i}ve/immature T cells, cytotoxic T cells, helper T cells, regulatory T cells, tumor-associated macrophages, and dendritic cells--and tumor subpopulations with varying immunogenicity. Additionally, the model incorporates an intercellular communication network mediated by key cytokines, including IL-2, IL-10, IL-12, IFN-$\gamma$, and TGF-$\beta$.
	
The optimized QCIC model dynamically simulates key biological processes within the cancer-immunity cycle. First, apoptotic tumor cells in the tumor microenvironment release specific antigens, which are captured by resident dendritic cells. These antigen-laden cells migrate via the lymphatic network to tumor-draining lymph nodes. Simultaneously, na\"{i}ve or immature immune cells from the hematopoietic system (bone marrow and thymus) enter the lymph nodes via peripheral circulation. Within the lymph node microenvironment, antigen presentation induces the differentiation of na\"{i}ve or immature T cells into tumor-specific effector T cells. Finally, these activated effector cells home to the tumor site via the circulatory system to exert specific cytotoxic effects. This multi-step, multi-organ framework effectively reproduces the physiological immune surveillance mechanism and establishes a quantitative prediction system spanning molecular to organ levels, providing an innovative systems biology platform for evaluating RCC immunotherapy.
	
In this study, we established a framework to evaluate treatment efficacy based on tumor volume dynamics, quantifying tumor burden before and after treatment. The QCIC model demonstrates robust clinical predictive capability; its simulation results align closely with observational data from real-world patient cohorts, enabling accurate characterization of individual treatment responses. This modelling strategy aims to guide clinical selection of optimal treatment pathways and, ultimately, to improve survival outcomes for patients with advanced RCC.
	
\section{Mathematical model}
\label{Sec2}
	
\subsection{QCIC modeling framework}
	
\begin{figure}[htbp]
\centering
\includegraphics[width=15.5cm]{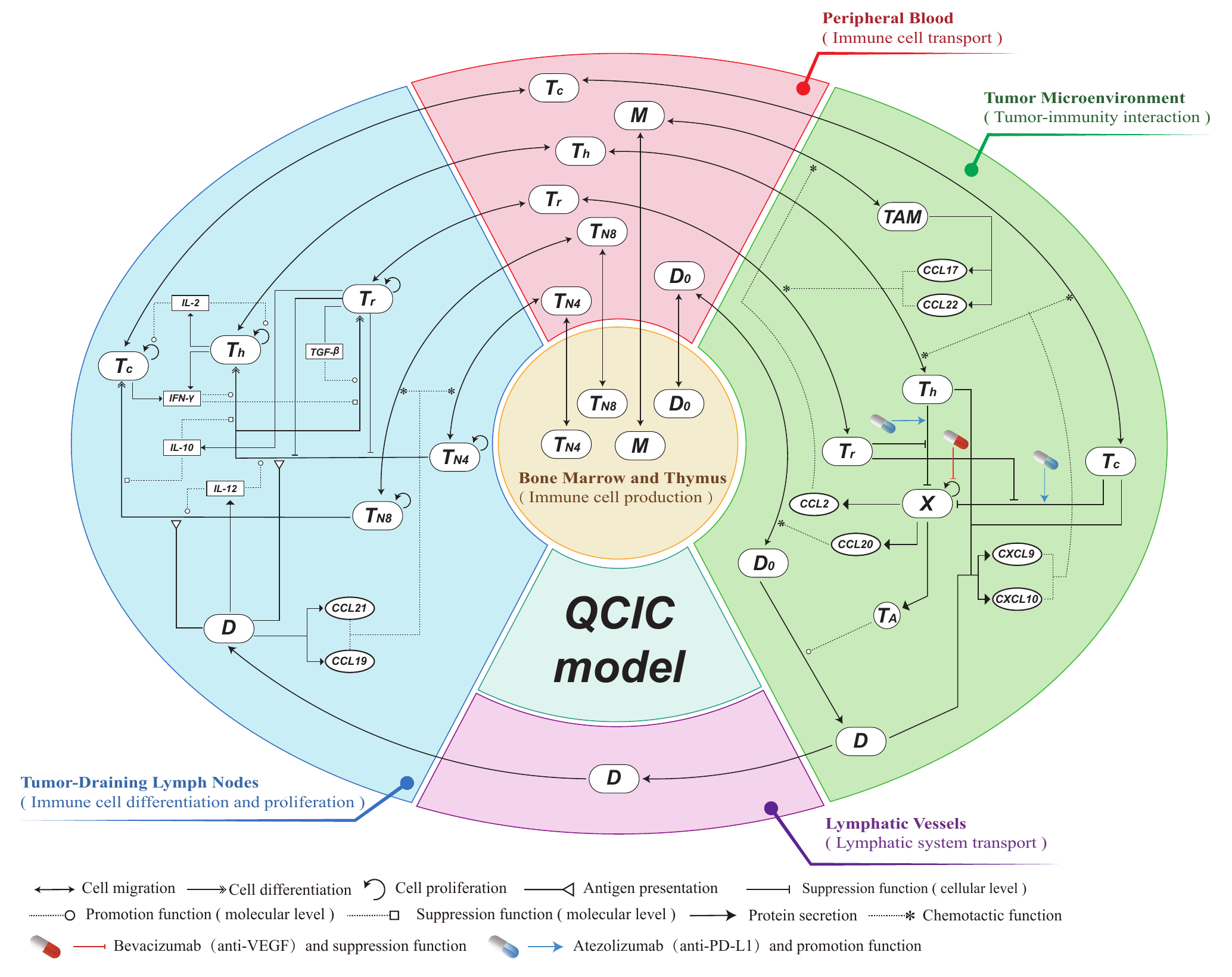}
\caption{{\bf QCIC model schematic.} The model consists of bone marrow and thymus (compartment $\mathbf{A}$), peripheral blood (compartment $\mathbf{B}$), tumor-draining lymph nodes (compartment $\mathbf{C}$), tumor microenvironment (compartment $\mathbf{D}$), and lymphatic vessels (compartment $\mathbf{E}$).}
\label{fig:1}
\end{figure}

Based on the theoretical framework of the tumor-immunity cycle proposed by Mellman et al. \citep{Mellman.Immunity.2013,Mellman.Immunity.2023}, and the QCIC model established in our previous study \citep{Li.NPJSystBiolAppl.2025}, we developed a five-compartment system dynamics model (Fig. \ref{fig:1}). We varied the physiological system into five compartments: (1) the immune cell generation compartment (bone marrow and thymus), serving as the source for the generation and development of immature immune cells; (2) the circulatory system compartment (peripheral blood), responsible for immune cell transport and drug distribution; (3) the immune activation compartment (tumor-draining lymph nodes, TDLNs), where immature immune cells receive antigen stimulation and differentiate into effector cells; (4) the effector compartment (tumor microenvironment, TME), which serves as the primary site of tumor-immune interaction; and (5) the antigen transport compartment (lymphatic vessels), which mediates the transport of dendritic cells (DCs) and antigen information. 
	
The model dynamically simulates the entire tumor-immunity cycle: antigens released from apoptotic tumor cells in the TME are captured by DCs, which then migrate through lymphatic vessels to the draining lymph nodes. Concurrently, na\"{i}ve or immature T cells produced in the immune cell generation compartment reach the lymph nodes via peripheral blood, differentiate into effector cells under the influence of antigen presentation, and finally home to the tumor site through the blood circulation to exert specific cytotoxic effects. This modeling framework implements a systemic simulation ranging from immune cell generation and activation to effective killing, providing a new paradigm for the quantitative study of tumor immunotherapy. The key assumptions for model construction are outlined below. The detailed mathematical formulations and variable descriptions are provided in Supplementary Text 1 and 2.
	
\subsubsection{Cellular dynamics}
	
In the QCIC modeling process, the dynamic behabior of cell type $i$ encompasses eight distinct processes: source ($\mathcal{S}_{i}$), differentiation ($\mathcal{D}_{i}$), proliferation ($\mathcal{P}_{i}$), transformation ($\mathcal{T}_{i}$), migration ($\mathcal{V}_{i}$), chemotaxis ($\mathcal{X}_{i}$), killing ($\mathcal{I}_{i}$), and apoptosis ($\mathcal{A}_{i}$). Accordingly, the dynamics of cell type $i$ in compartment $n$ as described as follows: 
\begin{equation}
\label{eq:i}	    
\frac{d[i]_{n}(t)}{dt} = \mathcal{S}_{i}+\mathcal{D}_{i}+\mathcal{P}_{i}+\mathcal{T}_{i}+\mathcal{V}_{i}+\mathcal{X}_{i}+\mathcal{I}_{i}+\mathcal{A}_{i},
\end{equation}
where $[i]_n$ represents the density of cell type $i$ in compartment $n$. Based on known biological mechanisms of tumor-immune interactions, we provide specific mathematical descriptions for these eight dynamic behaviors: 
	
\textbf{Cell source term (${\mathcal{S}} _{i}$):} Since na\"{i}ve or immature immune cells are continuously produced in the bone marrow or thymus, the influx rate is described as:
	
\begin{equation}
\label{eq:S_{i}}	    
\mathbf{\mathcal{S}}_{i} = \theta_{i}, 
\end{equation}
where $\theta_{i}$ represents the production rate of cell type $i$. 
	
\textbf{Cell differentiation term (${\mathcal{D}} _{i}$):} In TDLNs, na\"{i}ve T cells interact with antigen-presenting cells and differentiate into effector T cells with distinct biological functions under cytokine regulation. We describe the differentiation term for cell type $i$ as:
\begin{equation}
\label{eq:D_{i}}	
\mathbf{\mathcal{D}}_{i} = \kappa_{i}\left(1+\sum_{p}\lambda_{ip}\frac{[p]_{m}}{K_{p}+[p]_{m}}\right)\prod_{q}\frac{1}{1+{[q]_{m}}/{K_{iq}}}\frac{[D]_{m}^{n}}{[D]_{m}^{n}+K_{D}^{n}} [i_0]_{m}, 
\end{equation}
where $[p]_{m}$ and $[q]_{m}$ represent the concentrations of regulating cytokines $p$ (promoting) and $q$ (inhibiting) in compartment $m$, respectively; $[D]_{m}$ represents the density of activated DCs; and $[i_0]_m$ denotes the density of na\"{i}ve T cells. Michaelis-Menten kinetics are used to model cytokine regulation: $\kappa_i \left(1+\sum_{p}\lambda_{ip}\frac{[p]_{m}}{K_{p}+[p]_{m}} \right)$ denotes the promotion of differentiation by cytokine $p$, where $\kappa_{i}$ is the differentiation rate, $\lambda_{ip}$ is the enhancement coefficient, and $K_p$ is the half-saturation constant. The term $\prod_{q}\frac{1}{1+{[q]_{m}}/{K_{iq}}}$ represents inhibition by cytokine $q$, where $K_{iq}$ is the half-saturation constant. A Hill function $\frac{[D]_{m}^{n}}{[D]_{m}^{n}+K_{D}^{n}} [i]_{m}$ characterizes the antigen presentation process, describing the effective contact between mature DCs and na\"{i}ve T cells, where $K_D$ is the half-saturation constant and $n$ is the Hill coefficient.
	
\textbf{Cell proliferation term (${\mathcal{P}} _{i}$):} The proliferation logic varies by cell tyoe. First, a logistic model characterizes the proliferation of na\"{i}ve T cells:
\begin{equation}
\label{eq:P_{i1}}
\mathbf{\mathcal{P}}_{i} = \beta_{i}\left(1-\frac{[i]_{m}}{G_{i}}\right)[i]_{m}, 
\end{equation}
where $\beta_{i}$ is the proliferation rate and $G_{i}$ is the carrying capacity.
	
Second, the proliferation of effector T cells is regulated by the local cytokine environment, modeled as 
\begin{equation}
\label{eq:P_{i2}}
\mathbf{\mathcal{P}}_{i} = \beta_{i}\left(1+\sum_{p}\lambda_{ip} \frac{[p]_{m}}{K_{p}+[p]_{m}}\right)[i]_{m}, 
\end{equation}
where $\lambda_{ip}$ represents the enhancement coefficient of cytokine $p$.
	
Finally, we use a competitive logistic model to characterize the competition among tumor subpopulations. Considering four variants---drug-sensitive tumor cells (DSTC), atezolizumab-resistant tumor cells (ARTC), bevacizumab-resistant tumor cells (BRTC), and double-resistant tumor cells  (DRTC)---the proliferation is described as:	
\begin{equation}
\label{eq:P_{i3}}
\mathbf{\mathcal{P}}_{i} = \frac{\beta_{i}}{1+\delta \Lambda_{B}\frac{[B]_m^n}{K_B^n+[B]_m^n}} \left(1-\frac{1}{G_i}\sum_{j} a_{ij}[j]_{m}\right)[i]_{m}, 
\end{equation}
where $i$, $j$ denote tumor cell types. $\delta$ discriminates bevacizumab sensitivity ($\delta = 1$ for sensitive cells, $\delta = 0$ for resistant cells). $\Lambda_{B}$ represents the maximum efficacy of bevacizumab, $[B]_{m}$ is the drug concentrations of bevacizumab, $K_B$ is the half-saturation constant, and $a_{ij}$ represents the intraspecific/interspecific competition coefficients.
	
\textbf{Cell type transition term (${\mathcal{T}} _{i}$):} Tumor cells and immature antigen-presenting cells exhibit plasticity and may undergo phenotypic transitions under drug or antigen influence. Using a Hill function to represent these influences:
\begin{equation}
\label{eq:T_{i}}
\mathcal{T}_{i} = \rho_{ij}\frac{[p]_{m}^n}{K_{p}^n+[p]^n_{m}}[j]_{m}
			\ \mathrm{or} \
			\mathcal{T}_{j} = -\rho_{ij}\frac{[p]_{m}^n}{K_{p}^n+[p]_{m}^n}[j]_{m}, 
\end{equation}
where $\rho_{ij}$ is the transition rate from cell type $j$ to $i$, and $[p]_{m}$ is the concentration of the inducing agent.  
	
\textbf{Cell migration term (${\mathcal{V}} _{i}$):} Driven by concentration gradients and physiological flow, cells migrate between compartments:
\begin{equation}
\label{eq:V_{i}}
\mathcal{V}_{i}^{n} = \upsilon_{input}^{n}-\upsilon_{output}^{n} =\sum_{m}\upsilon_{i}^{mn}\frac{V_{m}}{V_{n}}[i]_{m}-\sum_{l}\upsilon_{i}^{nl}[i]_{n}. 
\end{equation}
Here, $\upsilon_{input}^{n}$ and $\upsilon_{output}^{n}$ represent influx and efflux, respectively. $\upsilon_{i}^{mn}$ is the migration rate from compartment $m$ to $n$, and $V_{m}, V_{n}$ denote compartment volumes.
	
\textbf{Cell chemotaxis term (${\Large \mathcal{X}} _{i}$):} Directed migration driven by chemokine gradients is modeled using Michaelis-Menten kinetics:
\begin{equation}
\label{eq:X_{i}}
\mathcal{X}_{i}^{n} = -\sum_{j}\chi_{i j}^{n m} \frac{[j]_{m}}{K_{j}+[j]_{m}}[i]_{n}
			\ \mathrm{or} \
			\mathcal{X}_{i}^{m} = \sum_{j}\chi_{i j}^{n m} \frac{[j]_{m}}{K_{j}+[j]_{m}}[i]_{n}, 
\end{equation}
where $\chi_{i j}^{nm}$ is the maximum chemotactic rate of cell $i$ from compartment $n$ to $m$ induce by chemokine $j$.
	
\textbf{Cell killing term (${\mathcal{I}} _{i}$)} Tumor cytotoxicity is primarily mediated by lymphocytes, enhanced by immune checkpoint inhibitors. The killing term is described as:	
\begin{equation}
\label{eq:I_{i1}}
\mathcal{I} _{i} = - \sum_{j}\eta_{j}[j]_{m}\left(1+\delta \Lambda_{A}\frac{[A]_m^{n}}{K_I^{n}+[A]_m^{n}}\right)\times (\frac{1}{1+[p]_m/K_{p}})[i]_{m},
\end{equation}
where $\eta_{j}$ is the killing rate by immune cell $j$. $\Lambda_{A}$ is the efficacy coefficient of atezolizumab, and $\delta$ indicates sensitivity. $[A]_{m}$ is the atezolizumab concentration. The term $\frac{1}{1 + [p]_m/K_{p}}$ represents inhibition nby regulatory cells (where $p$ specifically denotes Tregs).
	
\textbf{Cell apoptosis term (${\Large \mathcal{A}} _{i}$):} Natural cell death is modeld as:	
\begin{equation}
\label{eq:A_{i1}}
\mathcal{A}_{i} = -d_{i}[i]_{m}, 
\end{equation}
where $d_{i}$ is the apoptosis rate.
	
\subsubsection{Molecular level dynamics}
	
Given that cytokines do not cross membranes in our model, we focus on their production and degradation. Assuming molecular dynamics occur on a much faster time scale than cellular evolution, the dynamics of cytokine/chemokine $p$ in compartment $n$ are limited to:
\begin{equation}
\label{eq:C_{i}}
\frac{d[p]_{n}(t)}{dt} = \mathcal{G}_{p}+\mathcal{H}_{p}, 
\end{equation}
where $\mathcal{G}_{p} = \sum_{i=1}^{k}\delta_{pi}[i]_{n}$ represents production (with $\delta_{pi}$ being the secretion rate by cell type $i$), and $\mathcal{H}_{p} = -d_{p}[p]_{n}$ represents degradation.
	
\subsection{Parameter estimation}
	
To accurately characterize the biological features of the tumor-immunity cycle, the QCIC model integrates key processes including cell proliferation, differentiation, transformation, migration, chemotaxis, killing, apoptosis, and cytokine dynamics. The system comprises 164 parameters, determined through a three-step calibration process: (1) fixing the order of magnitude for parameters based on literature; (2) constraining ranges for undocumented parameters based on biological plausibility; and (3) optimizing parameters via calibration against clinical data to ensure consistency with real-world observations. Sensitivity analysis was subsequently performed to identify parameters significantly impacting model output, which were then used to generate a virtual patient cohort for treatment optimization.
	
Parameter ranges were defined as follows: Cell mobility ($v_i^{mn}$) ranges from $10^ {-3}$ to $10^1\ \mathrm{day} ^ {-1}$ \citep{Wang.JITC.2021,Li.Heliyon.2022,Vibert.PLoSComputBiol.2017}. Compartment volume ($V_m$) range from $10^ {-2}$ to $10^1 \ \mathrm{L}$ \citep{Li.Heliyon.2022}. Chemotaxis rates ($\chi^ {mn}_{ip}$) range from $10^ {-2}$ to $10^ {-1} \ \mathrm{day} ^ {-1}$ \citep{Gunn.PNAS.1998,Loetscher.JEM.1996,Taub.JEM.1993,Iellem.JEM.2001,Bleul.JEM.1996}. Based on previous research \citep{Li.Heliyon.2022,Edd.iScience.2022,Li.BullMathBiol.2024,Li.NPJSystBiolAppl.2025}, the production rate of immune cells ($\theta_{i}$) is set between $10^{6}$ and $10^9\ \mathrm{cell} \cdot \mathrm{L} ^{-1} \cdot \ \mathrm{day}^{-1} $. Apoptosis rates ($d_i$) for cells ranged from $0.05$ to $0.1 \ \mathrm{day}^ {-1}$ \citep{Lai.PNAS.2018,Li.Heliyon.2022,Hao.PLoSOne.2016}, and the degradation rates of cytokines ($d_j$) ranged from $10^ {1}$ to $10^2\ \mathrm{day}^ {-1}$ \citep{Lai.PNAS.2018,Wang.JITC.2021,Hao.PLoSOne.2016,Zandarashvili.JBiolChem.2013,Lai.BMCSystemsBiology.2017,Chen.BMB.2012,Szomolay.JTB.2012,Lambeir.JBiolChem.2001}. The half-saturation constants for cytokines ($K_j$) range from $10^{3}$ to $10^{7}\ \mathrm{pg} \cdot \mathrm{L} ^ {-1}$, and the secretion rates ($\delta_{ji}$) of cytokines from $10^{-6}$ to $10^1 \ \mathrm{pg} \cdot \mathrm{cell} ^{-1} \cdot \mathrm{day} ^{-1}$ \citep{Li.FrontImmunol.2023,Lai.PNAS.2018,Lai.SciChinaMath.2020}. The Hill coefficient $n$ typically ranges from $1$ to $10$ \citep{Li.BullMathBiol.2024,Lai.PNAS.2018}.The immune killing rate ($\eta_{j}$) ranges from $10^{-11}$ to $10^ {-10} \ \mathrm{day} ^ {-1}$. Detailed parameter values and sources are provided in Supplementary Text 3 and Supplementary Tables  S8--S17.
	
\section{Results}
\label{sec3}
	
\begin{figure}[htbp]
		\centering
		\includegraphics[width=15.5cm]{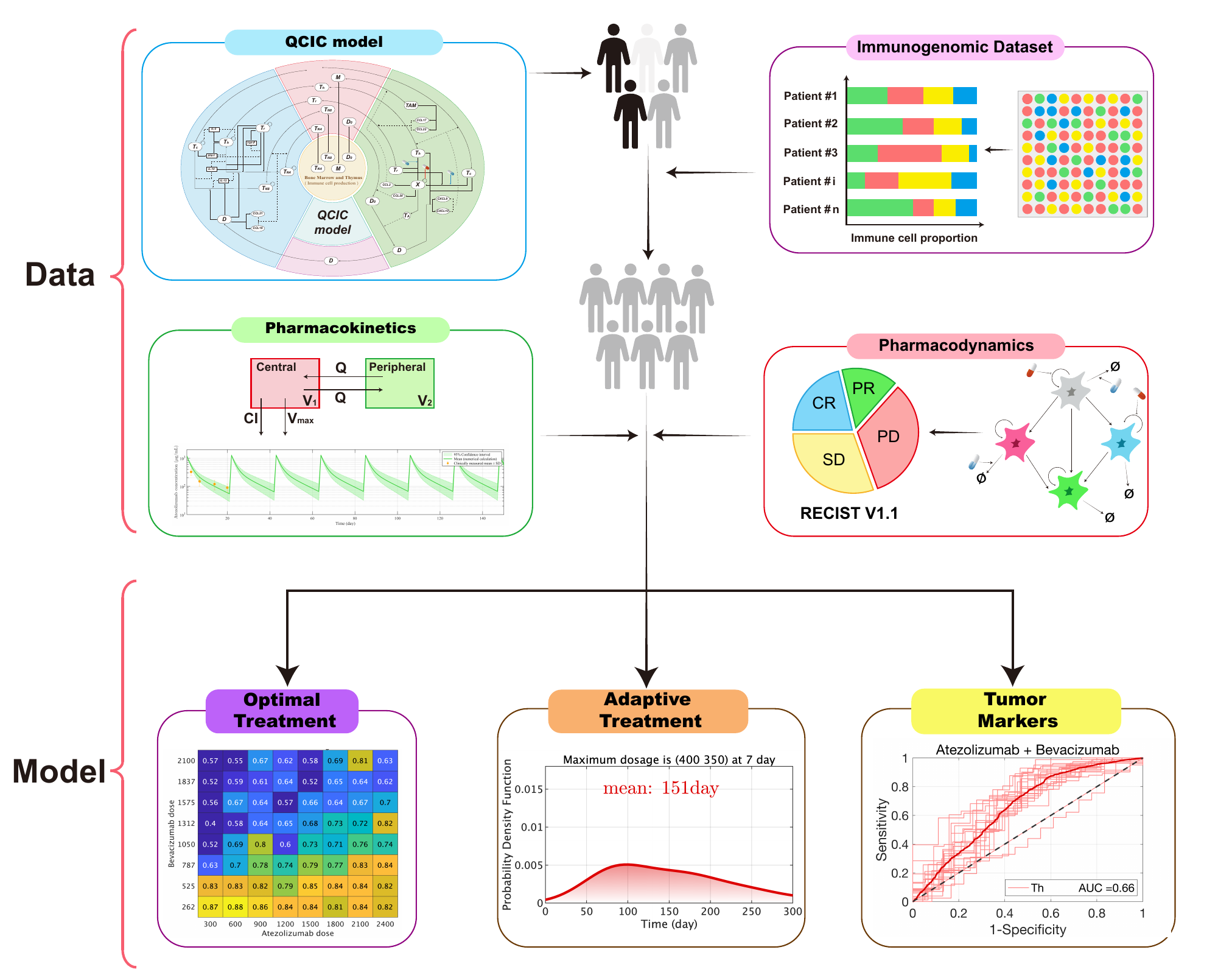}
		\caption{{\bf Schematic workflow of the data- and model-driven QCIC framework.} This study proceeds from model establishment to clinical application. First, initial parameter sets (indicated by colored silhouettes) are generated via Beta distribution sampling and refined through calibration with clinical and experimental data to construct a biologically valid virtual patient cohort. Next, a pharmacokinetic (PK) model is integrated to simulate drug disposition. Virtual patients are then stratified into distinct response subgroups based on heterogeneity in treatment outcomes. Finally, this platform is applied to optimize therapy by comparing fixed-dose versus adaptive administration strategies and to evaluate dynamic tumor biomarkers. RECIST v1.1: Response Evaluation Criteria in Solid Tumors 1.1}
		\label{fig:2}
	\end{figure}

Figure \ref{fig:2} illustrates the comprehensive workflow of this study. In the model construction phase, we employed an enhanced Quantitative Cancer-Immunity Cycle (QCIC) model to generate a parameter set via beta distribution sampling. These parameters were subsequently calibrated using clinical and experimental data from patients with renal cell carcinoma (RCC) (Section \ref{sec3-1}). Detailed parameter interpretation and values are provided in Supplementary Text 3. Furthermore, we integrated pharmacokinetic analysis (Section \ref{sec3-2}) to refine the model framework. Based on drug response heterogeneity, patients were stratified into subgroups at both population and individual levels (Section \ref{sec3-3}).
	
In the application phase, we utilized the generated virtual patient cohort to compare the therapeutic efficacy of fixed-dose administration (Section \ref{sec3-4}) versus adaptive dosing strategies (Section \ref{sec3-5}), aiming to optimize personalized treatment plans. Finally, we evaluated the dynamic evolution of tumor biomarkers under combination therapy (Section \ref{sec3-6}).
	
\subsection{Virtual patient cohort for advanced RCC and verification of immune characteristics}
\label{sec3-1}
	
To characterize the population-level immunological heterogeneity of patients with advanced RCC, we integrated a stochastic modelling approach, immunogenomic data, and the QCIC model to recapitulate the immune microenvironment. This establishes a reliable virtual patient platform for the subsequent evaluation of various immunotherapeutic strategies. The specific steps are outlined below:
	
\textbf{(1) Sensitivity analysis of immune heterogeneity parameters.} Based on the clinical immunological characteristics of real patients, we identified three key categories of parameters to characterize inter-patient immune response differences: the immune cell production rates $\theta_i = { \theta_{D_0}, \theta_{T_{N4}}, \theta_{T_{N8}}, \theta_{M} }$, differentiation rates $\kappa_i = { \kappa_{T_h}, \kappa_{T_r}, \kappa_{T_c} }$, and proliferation rates $\beta_i = { \beta_{T_h}, \beta_{T_r}, \beta_{T_c} }$. To further analyze the importance of these selected parameters, we performed robustness and sensitivity analyses. We determined the relative robustness index, $RRI= {M_{ij}}/{M_{ij}^*}$, to study parameter robustness, where $M_{ij}$ represents the change in immune cell density $j\ ([T_h]_D, [T_r]_D, [T_c]_D, [TAM]_D)$ induced by a change in the parameter $i\ (\theta_i, \kappa_i, \beta_i )$, and $M_{ij}^*$ denotes the baseline value. We then employed least squares to fit a linear regression $\vec{J} = \vec{\omega } \times \vec{I}$, where $\vert \vec{\omega} \vert$ quantifies the sensitivity of immune cell $j$ to parameter $i$.
	
\begin{figure}[htbp]
		\centering
		\includegraphics[width=15.5cm]{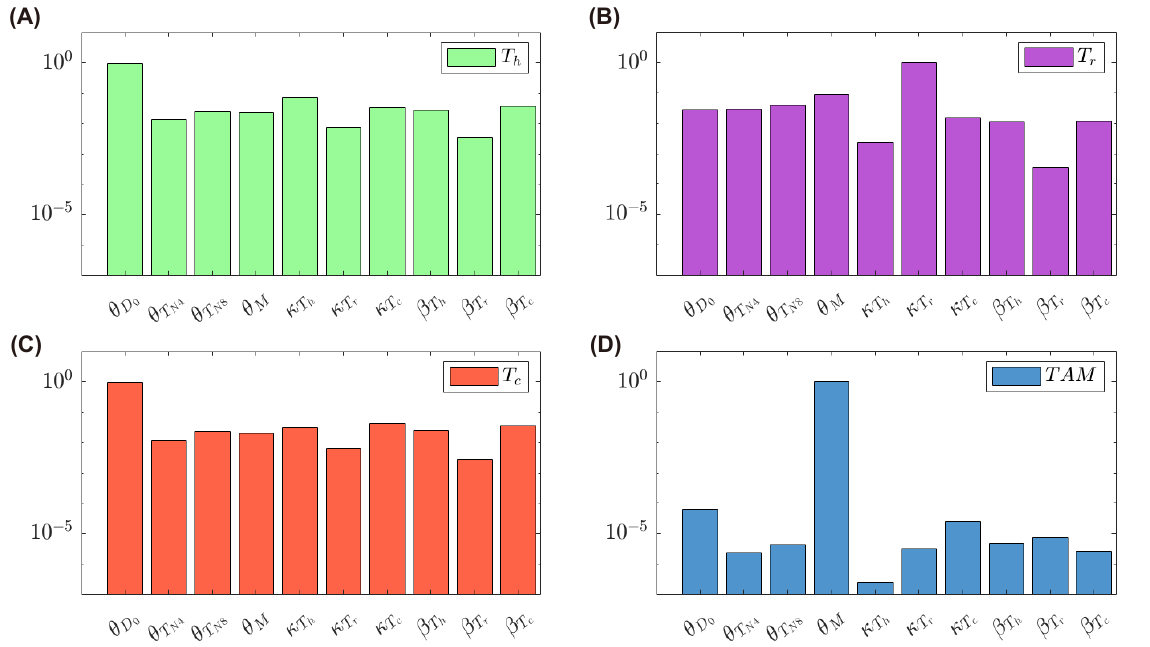}
		\caption{{\bf Parameter sensitivity analysis.} Sensitivity analysis of the tabulated parameters on immune cells, including $T_h $ (A), $T_r $ (B), $T_C$ (C), and $TAM $ (D), plotted on a logarithmic axis at day 50.}
		\label{fig:3}
	\end{figure}

The results indicate that perturbations in $\theta_{T_{N4}}$ have a significant effect on late-stage immune cells $[T_h]_D $, $[T_c]_D $, and $[T_r]_D $. Perturbation in $\beta_ {i}$ and $\kappa_ {i}$ $(i={T_h, T_c})$ significantly impact their corresponding cell types. $\theta_{M}$ significantly affects $[TAM]_D$. Furthermore, perturbations in $\beta_{T_{r}} $, $\kappa_{T_{r}} $, and $\theta_{D_{0}}$ significantly influence $[T_h]_D$, $[T_r]_D$, and $[T_c]_D$, as shown in Supplementary Figure S1. For early-stage immune dynamics, $\theta_{D_0}$ exhibits the highest sensitivity toward $[T_h]_D$ and $[T_c]_D$, $\kappa_{T_{r}}$ toward $[T_r]_D$, and $\theta_{M}$ toward $[TAM]_D$, as illustrated in Fig. \ref{fig:3} (A--D). These results confirm that our chosen parameters can effectively represent the heterogeneity of a virtual patient.
	
\textbf{(2) Modelling individual differences and initial states.} For each parameter $y$ in the selected set of immune heterogeneity parameters, we utilized the Beta distribution to randomly sample within the physiological range $[y_ \mathrm{min}, y_ \mathrm{max}]$. Specifically, parameters were extracted using the formula $y=y_ \mathrm{min}+(y_ \mathrm{max}-y_\mathrm{min}) \times x$, where $x \in (0,1)$ is a random variable following a Beta distribution with the probability density function:
$$
f(x;a,b) = \frac{x^{a-1}(1-x)^{b-1}}{B(a,b)}, \quad B(a,b)=\frac{\Gamma(a+b)}{\Gamma(a)\Gamma(b)},
$$
Here, $a$ and $b$ represent the shape parameters of the Beta distribution, determined by aligning model outputs with sequencing data. For each virtual patient, we randomly generated an initial tumor burden corresponding to the clinical range. We then simulated the natural tumor growth process over one year under untreated conditions. The simulated tumor burden at the endpoint was archived as a pre-processed benchmarking dataset. The physiological ranges and descriptions of the selected parameters are detailed in Table \ref{tab:1}.
	
	{\footnotesize{
	\begin{longtable}{c|c|c|c|c}
		\caption{Key immune heterogeneity parameters}
		\label{tab:1} \\
		\hline
		\toprule
		\footnotesize
		\textbf{Parameter} & \textbf{Description}
		&\textbf{Range}  & \textbf{Unit} & \textbf{Estimation*}  \\
		\midrule     
		\endfirsthead
		
		$\theta_{D_{0}}$ & Production rate of immature DCs & $1 \times 10^{7}$ -- $1 \times 10^{9}$ & $cell \cdot L^{-1} \cdot day^{-1}$ & 1, 2  \\
		
		$\theta_{T_{N4}}$ & Production rate of na\"{i}ve CD4+T cells & $5\times 10^{6}$ -- $8 \times 10^{9}$ & $cell \cdot L^{-1} \cdot day^{-1}$ & 1, 2   \\
		
		$\theta_{T_{N8}}$ & Production rate of na\"{i}ve CD8+T cells & $5 \times 10^{6}$ -- $8 \times 10^{9}$ & $cell \cdot L^{-1} \cdot day^{-1}$ & 1, 2  \\
		
		$\theta_{M}$ & Production rate of macrophages & $ 1 \times 10^{7}$ -- $8 \times 10^{8}$ & $cell \cdot L^{-1} \cdot day^{-1}$ & 2, 3   \\ 	\hline 
		
		$\kappa_{T_{h}}$ & Differentiation rate of helper T cells  & 0.1 -- 1 & $day^{-1}$ & 2, 3 \\
		
		$\kappa_{T_{r}}$ & Differentiation rate of regulatory T cells  & 0.1 -- 1 & $day^{-1}$ & 2, 3, 4 \\
		
		$\kappa_{T_{c}}$ & Differentiation rate of cytotoxic T cells  & 0.1 -- 1 & $day^{-1}$ & 2, 3  \\ 	\hline 
		
		$\beta_{T_{h}}$ & Proliferation rate of helper T cells  & 0.2 -- 0.45 & $day^{-1}$  & 2,5 \\      
		
		$\beta_{T_{r}}$ & Proliferation rate of regulatory T cells  & 0.1 -- 0.6 & $day^{-1}$  & 2,5 \\         
		\tiny
		$\beta_{T_{c}}$ & Proliferation rate of cytotoxic T cells  & 0.10 -- 1 & $day^{-1}$  & 2, 5 \\
		
		\midrule 
	\end{longtable}
			
	\begin{minipage}{14cm}
	1 = \cite{Edd.iScience.2022}, 2 = \cite{Li.NPJSystBiolAppl.2025}, 3 = \cite{Li.Heliyon.2022}, 4 = \cite{Wang.iScience.2022}, 5 = \cite{Vibert.PLoSComputBiol.2017}
	\end{minipage}
}
	}\\
	
\textbf{(3) Construction of the virtual patient cohort.} We constructed a virtual patient cohort exhibiting population-level immunological diversity by generating 20,000 random samples based on the clinical distribution characteristics of the heterogeneous parameters. This cohort comprehensively covers the potential immunophenotypic variations found in patients with advanced RCC.
	
\textbf{(4) Selection and calibration of the virtual patient cohort.} Utilizing the probability inclusion method proposed by \citet{Allen.CPT.2016}, we analyzed and downloaded clinical immune cell proportion data (including CD8, CD4, Treg, macrophages) from the iAtlas portal. We cleaned the experimental data by excluding outliers with cell proportions less than $0.05$ and calculated three immune characteristics: CD8/CD4, CD4/Treg, and Treg/TAM ratios. Virtual patient screening was achieved by incorporating the following probability model \citep{Wang.NPJPrecisOncol.2023}.
	
\begin{equation}
\label{eq:1}
{P(S(\theta)=1 \vert M(\theta)=r) = \beta\frac{\rho_{obs}(r) }{\rho_{sim}(r)}}, 
\end{equation}
where the parameter set $\theta$ represents a single virtual patient. The logical function $S (\theta)$ indicates whether the virtual patient corresponding to $\theta$ is selected (if $S (\theta)=1$). $M(\theta)$ describes the predicted dynamic changes in immune cell proportions. 
	
In this study, we focused on the preprocessing ratios of model predictions corresponding to available CD8/CD4, CD4/Treg, and Treg/TAM data. The key screening probability $P(S (\theta)=1 \vert M (\theta)=r)$ is determined by the ratio of the probability density of observed data to simulated data, $\rho_ {obs}(r)/ \rho_ {sim}(r)$. Density estimation uses the k-nearest neighbour algorithm: $\rho(r)=\frac {N}{V_N(r)} $, where $V_N(r)$ is the volume of an $N$-dimensional hypersphere with a radius from the target point $r$ to its $N$-th nearest neighbor ($N$ is typically chosen between $5$ and $10$). By optimizing the coefficient $\beta$ via a simulated annealing algorithm, the cumulative distribution difference between the virtual cohort and real patient data was minimized (Kolmogorov-Smirnov test), finally selecting the most clinically representative virtual patients (Fig. \ref{fig:4} A--C). The clinical data used to calibrate the model parameters are summarized in Supplementary Table 1.
	
To systematically evaluate the consistency between the virtual patient cohort and real clinical data, we quantified the distribution similarity of immune cell proportions using the Kullback-Leibler (KL) divergence and the Jensen-Shannon (JS) divergence:
\begin{equation}
\label{KL}
D_{KL}(P||Q)=\sum_{x}P(x)\mathrm{log}(\frac{P(x)}{Q(x)}),
\end{equation}
\begin{equation}
\label{JS}
D_{JS}(P||Q)=\frac{D_{KL}(P||M)+D_{KL}(Q||M)}{2},
\end{equation}
where $P(x)$ and $Q(x)$ represent the probability distributions of real and virtual patients, respectively, and $M(x)=\frac{1}{2} (P(x)+Q(x))$ is the mean distribution. The calculations show that the JS divergences for the immune characteristic combinations (CD8/CD4, CD4/Treg, and Treg/TAM) are $0.0086$, $0.0018$, and $0.0022$, respectively; the corresponding KL divergences are $0.0337$, $0.0072$, and $0.0085$ (Fig. \ref{fig:4} D--F). These results indicate that the virtual patient cohort constructed in this study not only accurately reproduces the compositional ratios of the four key immune cells (CD8+T, CD4+T, Treg, and TAM) but also reflects the characteristics of the tumor immune microenvironment. 
	
\begin{figure}[htbp]
\centering
\includegraphics[width=0.9\linewidth]{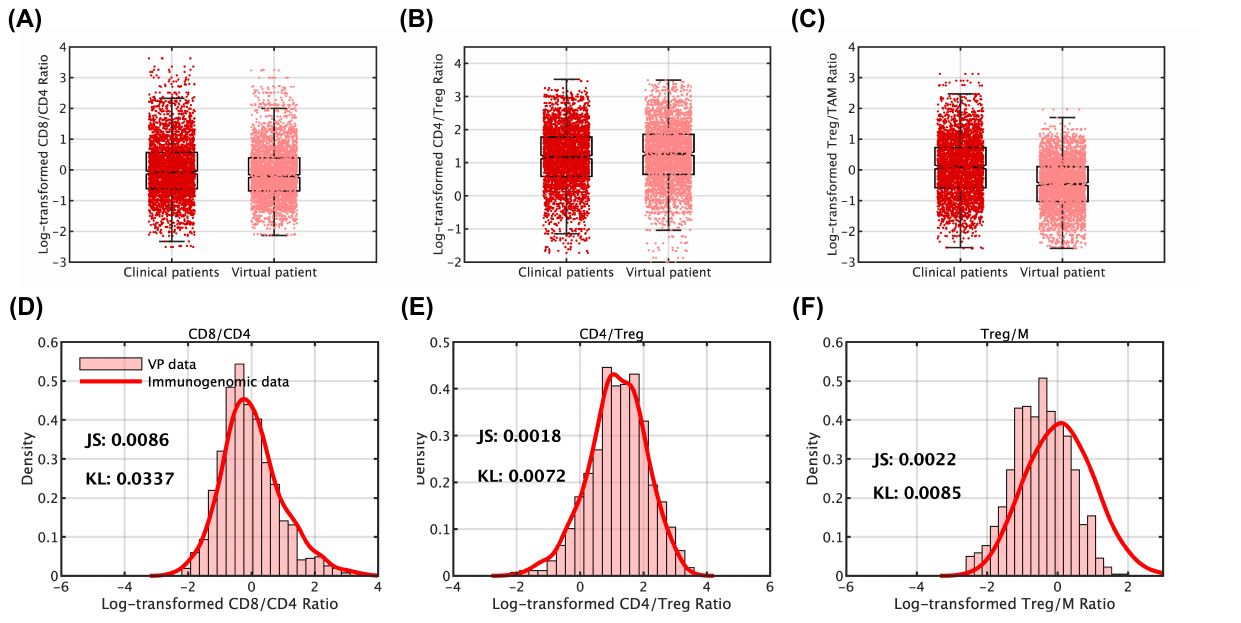}
\caption{{\bf Comparison of the virtual patient cohort and immunohistochemistry data.} {\bf (A--C)} Box plots. During virtual patient cohort generation, a logarithmic transformation was applied to immune cell subpopulations. Scatter points represent data from real patients and virtual patients, respectively. {\bf (D--F)} Probability density curve related to immunogenomic data (dark red) and data from randomly generated valid patients (pink bar chart). }
\label{fig:4}
\end{figure}
	
\subsection{Pharmacokinetic analysis of atezolizumab and bevacizumab}
\label{sec3-2}
	
Population pharmacokinetics (popPK) is an interdisciplinary approach merging classical compartmental modeling with statistical principles. By effectively integrating multi-source clinical data, popPK identifies typical population values, sampling ranges, and distribution characteristics of pharmacokinetic parameters. Given that anticancer drugs are administered at specific predetermined time points, we mathematically represented the dosing regimen $\hat{Q}(t)$ using the Dirac delta function $\delta$:
\begin{equation}
\label{eq:Drug} 
\hat{Q}(t)  = \sum_{i=1}^{n} D_{\mathrm{dose}} \cdot \delta(t-\tau_i),\quad i=1, 2, \dots, n,
\end{equation}
where $D_{\mathrm{dose}}$ represents the drug injection dose (amount) used in clinical trials, and $\tau_i$ denotes the administration time points. In this study, the standard dosing schedule for both atezolizumab and bevacizumab is every three weeks (Q3W). 

To accurately reflect clinical trial outcomes and characterize drug disposition within the body, we adopted a two-compartment pharmacokinetic model based on existing popPK studies. Since direct coupling between the selected immunogenomic data and PK parameters was unavailable, we independently generated PK parameters for virtual patients to match the heterogeneity observed in clinical cohorts. The dynamics of drug concentrations in the intervals between doses ($t \neq \tau_{i}$) are governed by the following system of ordinary differential equations (ODEs):
\begin{eqnarray}
\label{eq:Drug1} 
V_1 \frac{\mathrm{d} [\mathrm{Drug}]_1}{\mathrm{d} t} &=& Q \cdot ([\mathrm{Drug}]_2 - [ \mathrm{Drug}]_1) - \mathrm{CI} \cdot [\mathrm{Drug}]_1 - V_{\max} \frac{[\mathrm{Drug}]_1}{[ \mathrm{Drug}]_1 + K_M},\\
V_2 \frac{\mathrm{d} [\mathrm{Drug}]_2}{\mathrm{d} t} &=& Q \cdot ([\mathrm{Drug}]_1 - [\mathrm{Drug}]_2),
\end{eqnarray}
where $[\mathrm{Drug}] _1$ and $[\mathrm{Drug}] _2$ denote the drug concentrations in the central and peripheral compartments, respectively; $V_1$ and $V_2$ are the volumes of distribution; $Q$ is the inter-compartmental clearance; $\mathrm{CI}$ is the linear clearance rate; $V_ {\max}$ is the maximum non-linear clearance rate; and $K_M$ is the Michaelis-Menten constant.

To model the administration process, we implemented an instantaneous state update (impulse dosing) at each dosing time $\tau_i$. Specifically, the concentration in the central compartment is reset as follows:
\begin{equation}
\label{eq:DrugUpdate}
[\mathrm{Drug}]_1(\tau_i^+) = [\mathrm{Drug}]_1(\tau_i^-) + \frac{D_{\mathrm{dose}}}{V_1},
\end{equation}
where $[\mathrm{Drug}]_1(\tau_i^-)$ and $[\mathrm{Drug}]_1(\tau_i^+)$ represent the drug concentrations immediately before and after injection, respectively.

\begin{figure}[htbp]
\centering
\includegraphics[width=0.9\linewidth]{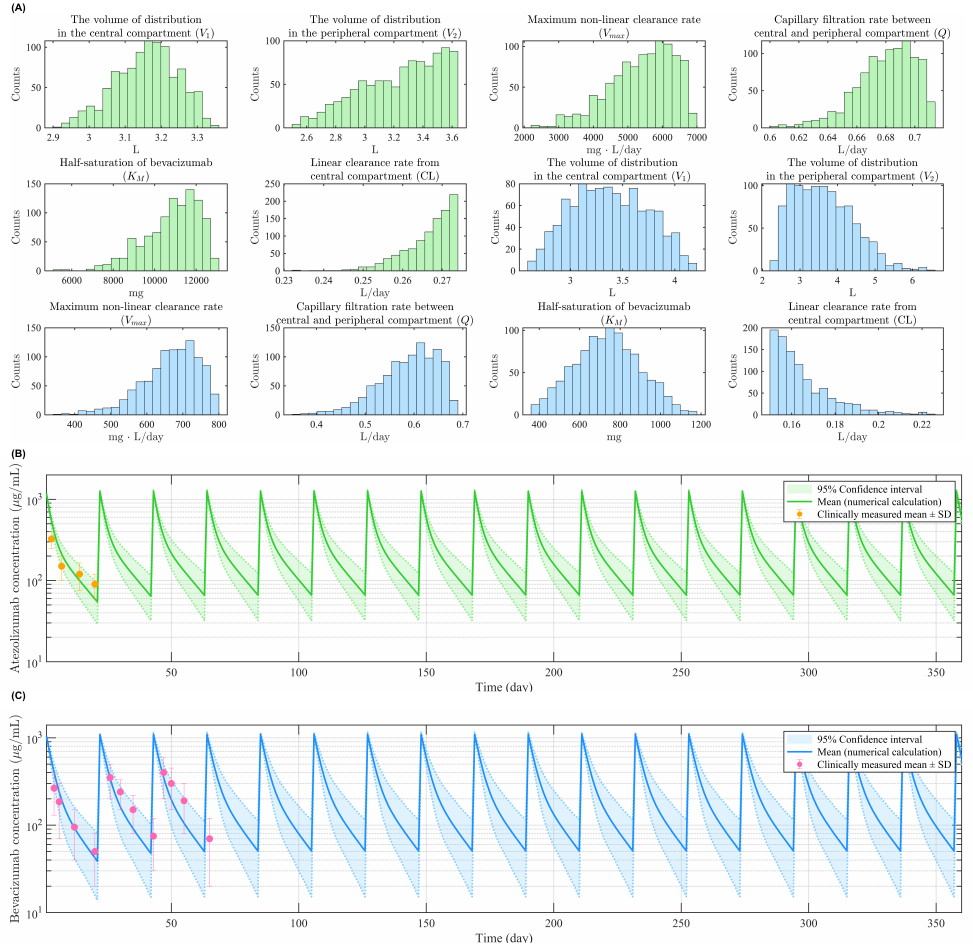}
\caption{{\bf Pharmacokinetic index of atezolizumab and bevacizumab.} {\bf (A)} The range and distribution of popPK parameters. {\bf (B)} Concentration-time profile of atezolizumab. The green solid line represents the mean plasma concentration, while the light green shaded area indicates the 95\% confidence interval. Orange markers represent clinically measured values.  {\bf (C)} Concentration-time profile of bevacizumab. The blue solid line represents the mean plasma concentration, while the light blue shaded area indicates the 95\% confidence interval. Purple markers represent clinically measured values.}
\label{fig:5}
\end{figure}

To capture inter-patient heterogeneity, we randomly sampled parameter sets from physiological ranges using a Beta distribution (Fig. \ref{fig:5}A). In the QCIC model, the drug concentration in the central compartment is assumed to represent the concentration within the tumor microenvironment. We simulated the dynamic changes in drug concentration over one year across different patients (Fig. \ref{fig:5}B--C). Detailed parameter ranges are provided in Table S4 of Supplementary Text 1.  Additionally, the dynamics of the peripheral compartment and the area under the curve (AUC) are shown in Figure S1 of Supplementary Text 1.
	
\subsection{Analysis of tumor progression at the population level}
\label{sec3-3}
	
To dynamically evaluate the therapeutic efficacy of treatments on tumors, clinical assessments typically utilize four categories: complete response (CR), partial response (PR), stable disease (SD), and progressive disease (PD). These criteria are defined based on changes in tumor burden relative to baseline. According to the RECIST v1.1 standard, CR is defined as a decrease of $\geq 80\%$ in tumor diameter, PR as a $\geq 30\%$ decrease in the sum of diameters, and PD as a $\geq 20\%$ increase; intermediate changes are classified as SD. To evaluate the performance of the QCIC model, we analyzed data from three clinical trials involving 1,031 patients with advanced RCC. Details on the processing of clinical data are provided in Supplementary Table 1.
	
To compare model predictions with clinical trial data, we introduced the Tumor Response Index (TRI) to quantify volumetric changes before and after treatment:
\begin{eqnarray}
\label{eq:TRI}
\mathrm{TRI} &=& \frac{V_{T_\mathrm{after}}}{V_{T_\mathrm{before}}} -1,\\
		V_T&=&\frac{V_\mathrm{tumor} \times [X]_D \times V_D }{V_E},
\end{eqnarray}
where $V_ {T_\mathrm{before}}$ and $V_ {T_\mathrm{after}}$ represent the tumor volume before and after treatment, respectively; $[X]_D$ is the tumor cell concentration; $V_D$ is the volume of the tumor microenvironment; $V_ \mathrm{tumor}$ is the volume of a single tumor cell ($2.572 \times 10^{-9} \mathrm{cm}^{3}/\mathrm{cell}$); and $V_E $ is the tumor volume fraction ($V_E=0.37$) \citep{Wang.JITC.2021,Wang.NPJPrecisOncol.2023}. Adapted from RECIST v1.1 for volumetric assessment, we defined the classification thresholds as follows: PD if $\mathrm{TRI} \ge 0.2$; SD if $-0.3 \le \mathrm{TRI} < 0.2$; PR is $-0.8 \le \mathrm{TRI} < -0.3$; and CR if $\mathrm{TRI} < -0.8$.
	
To test the predictive capability of the QCIC model for short-term outcomes, we simulated untreated tumor growth for 360 days to establish baseline conditions. The final immune cell counts were used as initial values for the virtual patients. Simultaneously, initial tumor diameters were sampled from the distribution observed in advanced RCC patients. We then integrated the calibrated PK parameters into the model to investigate the effects of different therapeutic regimes. The dosing schedule consisted of 1200 mg of atezolizumab and 1050 mg of bevacizumab administered every 21 days (Q3W).

\begin{figure}[htbp]
\centering
\includegraphics[width=0.9\linewidth]{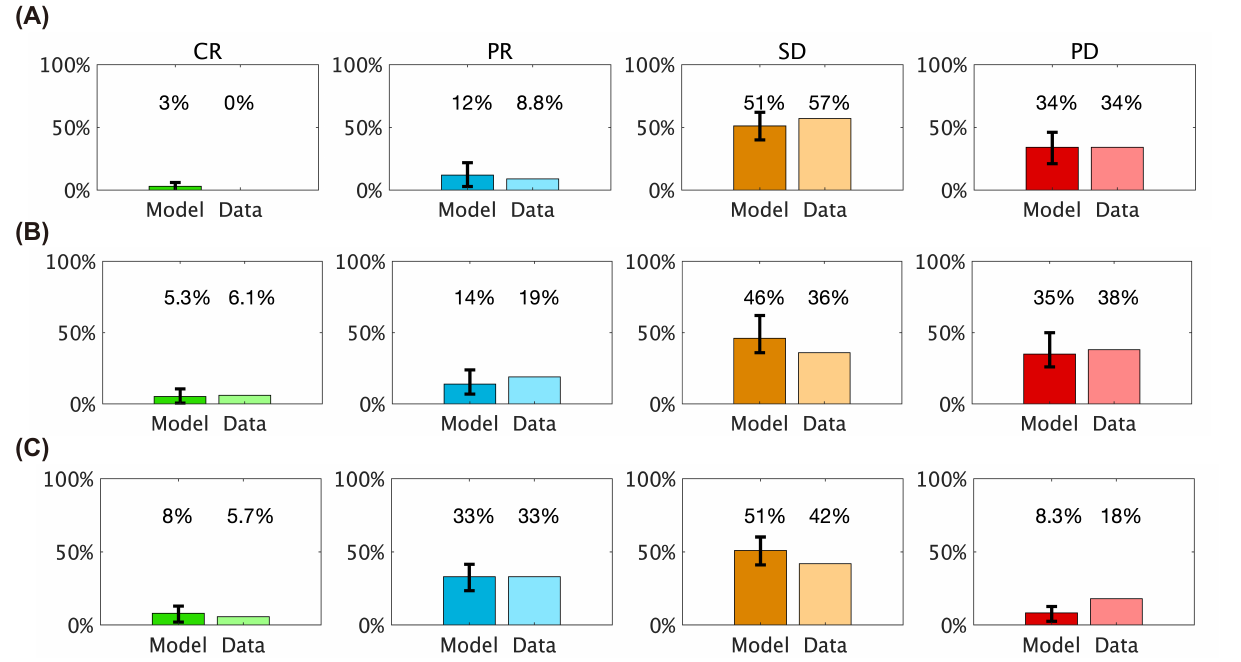}
\caption{{\bf Validation of the QCIC model against clinical tumor assessment criteria at the population level.} \textbf{(A)} Placebo group, data sourced from \cite{Motze.NEJM.2007}; \textbf{(B)} Atezolizumab treatment, data sourced from \cite{Rini.Lancecet.2019,Rini.Eur.2021}; \textbf{(C)} Combination treatment, data sourced from \cite{Rini.Lancecet.2019,Rini.Eur.2021}. Clinical simulation data is derived from 100 virtual patients randomly sampled from a cohort of 10,000 in the QCIC model. Solid tumor progression is evaluated based on the overall response after seven weeks of treatment. The RECIST calculation method for clinical data is derived from \cite{Li.NPJSystBiolAppl.2025}, while the RECIST  calculation for simulated data utilizes Eq. \ref{eq:TRI}. Error bars represent the $95\%$ confidence interval.}
\label{fig:6}
\end{figure}

From a generated cohort of 10,000 virtual patients per group, we randomly sampled 100 patients to calculate the average response rates. The simulation results indicated that in the placebo group, the rates of CR, PR, SD, and PD were $3\%$, $12\%$, $51\%$, and $34\%$, respectively, agreeing well with clinical values of $0\%$, $8.8\%$, $57\%$, and $34\%$ (Fig. \ref{fig:6}A). For atezolizumab monotherapy, the simulated rates were $5.3\%$ (CR), $14\%$ (PR), $46\%$ (SD), and $35\%$ (PD). While there was a slight deviation from clinical observations ($6.1\%$, $19\%$, $36\%$, and $38\%$, respectively), the results remained within the $95\%$ confidence interval (Fig. \ref {fig:6} B). In the combination therapy group, the simulated rates for CR, PR, SD, and PD were $8\%$, $33\%$, $51\%$, and $8.3\%$, respectively. Although the SD and PD rates varied from the clinical data ($5.7\%$, $33\%$, $42\%$, and $18\%$), the results fell within the confidence intervals. Notably, the overall non-response rates (SD + PD) were highly consistent at $59.3\%$ (simulated) and $60\%$ (clinical) (Fig. \ref{fig:6} C). These results demonstrate that the QCIC model effectively generates virtual patients that replicate clinical indicators for evaluating short-term treatment efficacy.
	
\begin{figure}[htbp]
\centering
\includegraphics[width=0.9\linewidth]{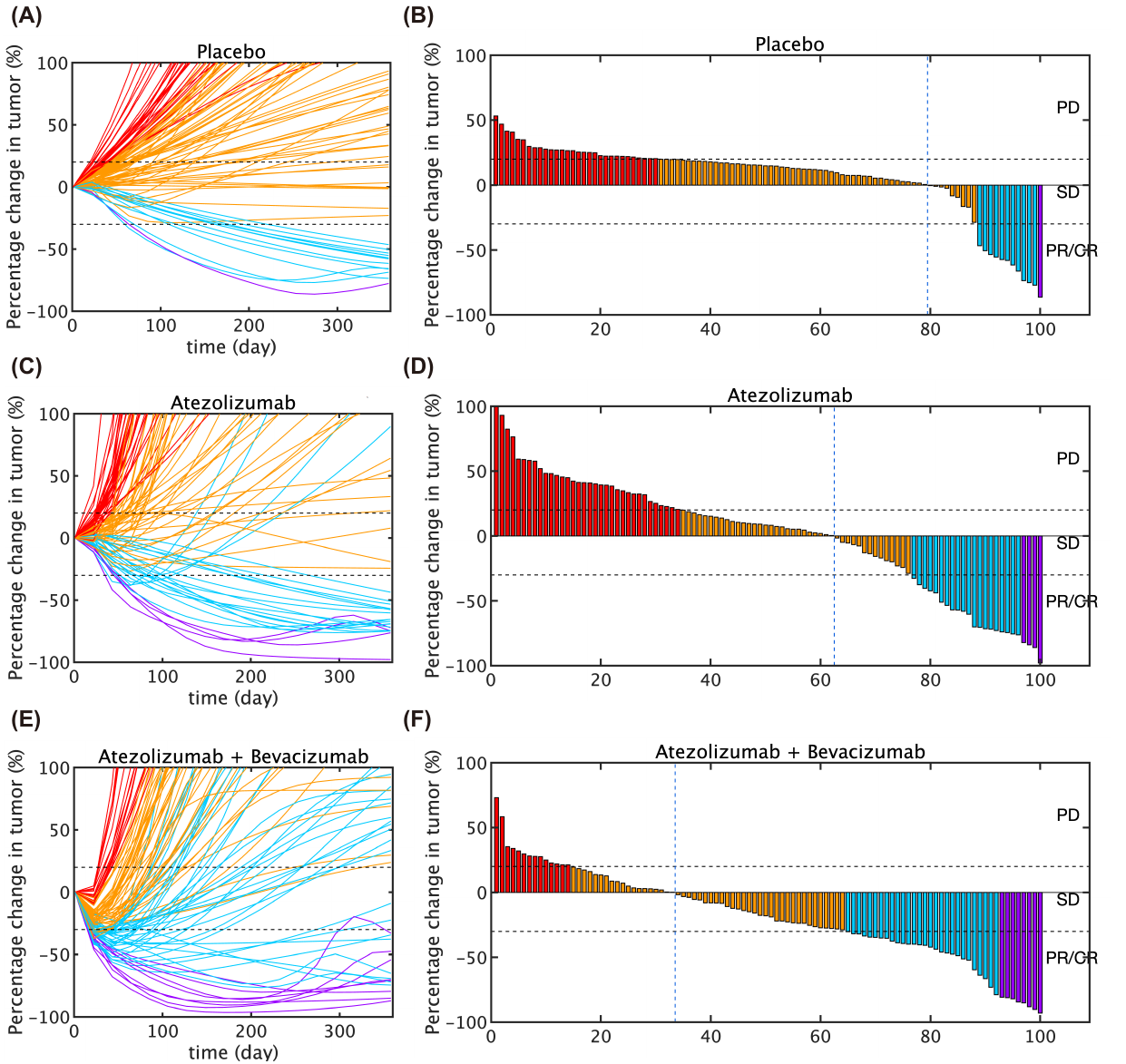}
\caption{{\bf The QCIC model simulates the dynamic evolution of tumors at the individual level.} Response kinetics (left) and overall response (right) of tumor diameters predicted by the QCIC model for 100 virtual patients. Changes in tumor diameter were assessed every three weeks according to tumor evaluation metrics in the placebo group \textbf{(A)}, the Atezoliauman group \textbf{(C)}, and the combined atelizumab $+$ bevacizumab group \textbf{(E)}. Each line represents a single virtual patient; black dotted lines indicate the thresholds for PD, SD, and PR/CR classification. In the waterfall plots (right) \textbf{(}\textbf{B}, \textbf{D}, \textbf{F}\textbf{)}, the blue dotted line represents tumor regression to the baseline level; bars to the left indicate regression (shrinkage), while the bars to the right indicate growth relative to baseline. Colors represent different responses: red (PD), orange (SD), blue (PR), and purple (CR).}
\label{fig:7}
\end{figure}

To further investigate the dynamic prediction capabilities of the QCIC model, we visualized the longitudinal treatment response of virtual patients using spider plots (left panels) and waterfall plots (right panels) every three weeks, as shown in Fig. \ref{fig:7}. For each group, we tracked 100 virtual patients to display changes in tumor diameter and overall response. Simulation results revealed that the proportion of tumors regressing to baseline levels ranged from $20.5\%$ in the placebo group to $37.5\%$ in the atezolizumab arm (Fig. \ref{fig:7}B, D), suggesting that approximately $17\%$ of patients derived specific benefit from atezolizumab. Furthermore, this proportion increased to $78.5\%$ in the combination therapy arm (Fig. \ref{fig:7}F), indicating that over $50\%$ of patients achieved significant clinical benefit compared to placebo. Compared to atezolizumab monotherapy, combination therapy provided effective treatment for an additional $40\%$ of patients, further highlighting the synergistic clinical benefits.
	
\subsection{Investigation of the effect of drug dosage on the recurrence rate of RCC under different treatment strategies}
\label{sec3-4}
	
\begin{figure}[htbp]
\centering
\includegraphics[width=0.9\linewidth]{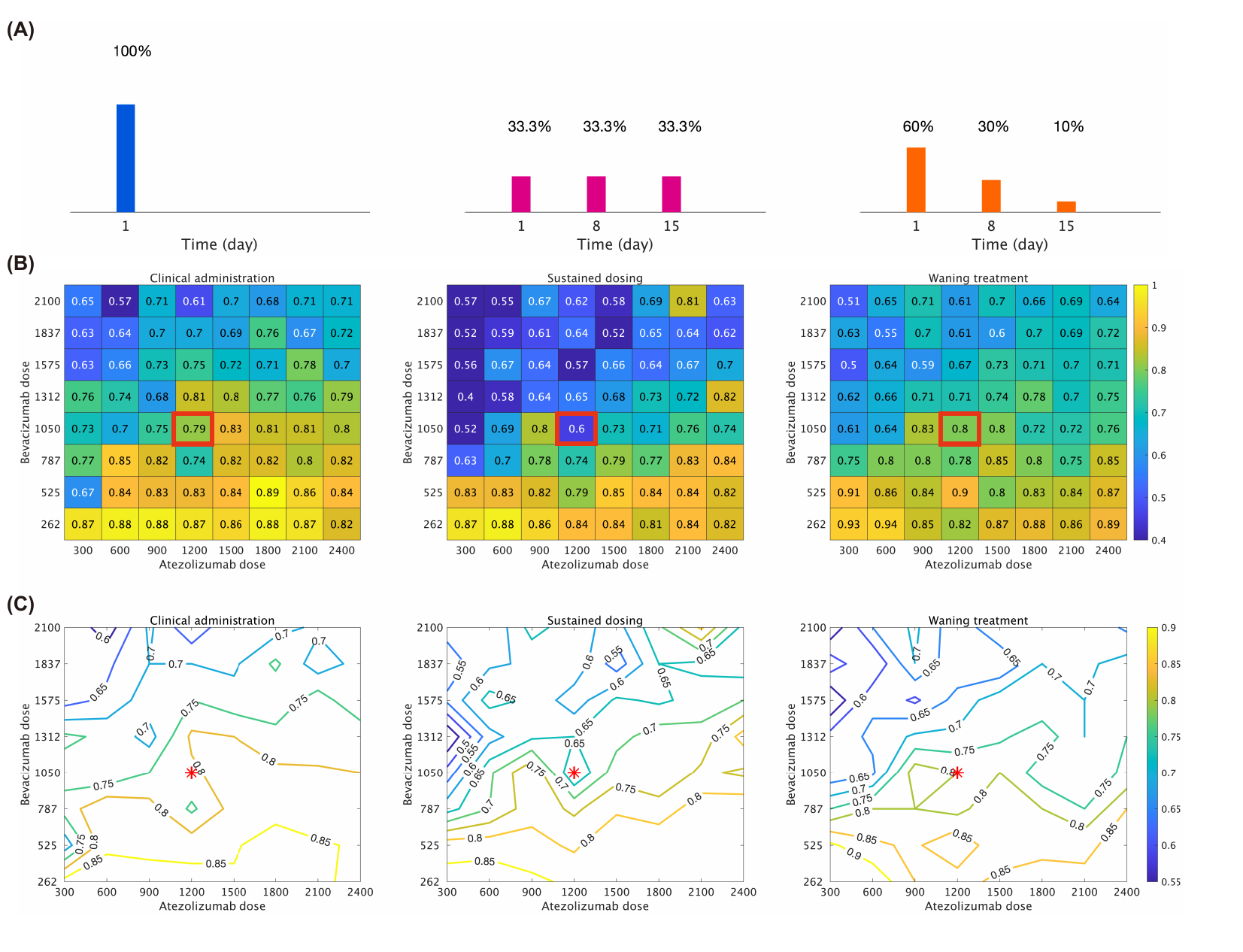}
\caption{{\bf Dose-response analysis of recurrence rates under differing scheduling strategies.} \textbf{(A)} Schematic illustration of the dosing schedules: clinical (Q3W), sustained (Q1W, constant dose), and waning (Q1W, tapered dose: $60\%$, $30\%$, and $10\%$). \textbf{(B)} Heatmaps visualizing predicted recurrence rates across the parameter space. Red squares indicate the standard dose $(1200, 1050)$. \textbf{(C)} Contour plots showing interpolated recurrence landscapes for all dose combinations. Red stars indicate the standard dose $(1200, 1050)$.}
\label{fig:8}
\end{figure}

To investigate the complex interplay between dosing schedule, drug dosage, and therapeutic outcome, we systematically evaluated three distinct temporal strategies (clinical, sustained, and waning administration) using a full factorial design. Drug doses were scaled from $0.25\times$ to $2.0\times$ the standard clinical dose of atezolizumab ($1200$ mg) and bevacizumab ($1050$ mg) (Fig. \ref{fig:8}A). The dose scaling factors applied were $0.25$, $0.5$, $0.75$, $1.0$, $1.25$, $1.5$, $1.75$, and $2.0$. Tumor recurrence was defined using the Tumor Response Index (Eq. \ref{eq:TRI}) with a threshold of $\mathrm{TRI} \ge 0.2$. Each combination was simulated in a cohort of $10,000$ virtual patients, and recurrence rates were calculated with $95\%$ confidence intervals based on $100$ iterative samplings (Fig. \ref{fig:8}).

First, we compared the three temporal strategies at the standard clinical dose $(1200, 1050)$ (Fig. \ref{fig:8}B, red squares). Sustained dosing achieved the lowest recurrence rate ($60\%$), which was $19\%$ lower than clinical administration ($79\%$) and $20\%$ lower than the waning treatment ($80\%$). This highlights the significant advantage of fractionating the total dose into smaller, more frequent administrations. In contrast, waning treatment resulted in a recurrence rate comparable to that of clinical administration, suggesting that maintaining bevacizumab exposure throughout the cycle--rather than tapering it off--is crucial for preventing recurrence. 

Next, simulation results across varying dosages revealed a pronounced synergistic interaction and a clear therapeutic hierarchy between the two agents. Across all three dosing schedules, bevacizumab emerged as the primary driver of efficacy. When bevacizumab doses were insufficient ($\le 1050$ mg), varying the atezolizumab dose had only a minor impact on reducing recurrence (Fig. \ref{fig:8}B). Conversely, increasing the bevacizumab dose consistently reduced recurrence rates, distinct from the independent effects of atezolizumab. Notably, the effectiveness of atezolizumab appeared conditional: significant reductions in recurrence were observed only when the bevacizumab dosage was sufficiently high ($\ge 1312$ mg). This suggests that anti-angiogenic therapy with bevacizumab primes the tumor microenvironment, enabling the subsequent efficacy for immune checkpoint inhibition.

Based on the contour plots in Fig. \ref{fig:8}C, we identified optimal dose combinations for each strategy. For clinical administration (Q3W), the regimen $(600, 2100)$--corresponding to low-dose atazolizumab and high-dose bevaciazumab--achieved the lowest recurrence rate ($57\%$), a $22\%$ reduction compared to the standard dose ($79\%$). For sustained dosing (Q1W), the optimal regimen was $(300, 1312)$, achieving a recurrence rate of approximately $40\%$, a further $20\%$ reduction compared to its standard dose. For waning treatment (Q1W, tapered), the optimal regimen was $(300, 1575)$, reducing the recurrence rate to $50\%$. Intriguingly, while simply doubling both doses $(2400, 2100)$ failed to improve outcomes in the clinical or sustained strategies (likely due to saturation or diminishing returns), it effectively controlled recurrence in the waning strategy.

Combining these findings, a detailed analysis of the prediction space highlights a critical optimal exposure window (Fig. \ref{fig:8}C, upper-left region). Most combinations achieving recurrence rates below $60\%$ involved bevacizumab doses at $1.25\times$ to $2.0\times$ the standard dose, paired with atezolizumab doses of only $0.25\times$ to $1.0\times$ the standard dose. This observation has significant implications for clinical translation, particularly regarding immunotherapy-related adverse events. If a patient exhibits intolerance to atezolizumab, switching from a clinical to a sustained or waning strategy is a promising alternative. These strategies allow for a significant reduction in atezolizumab dosage (e.g., to $0.25\times$ standard) while compensating with increased bevacizumab, ultimately achieving superior tumor control compared to the standard high-dose clinical regimen.

\subsection{Exploring patient recurrence under adaptive treatment strategies}
\label{sec3-5}
	
To further optimize therapeutic outcomes, we moved beyond fixed dosing schedules to explore adaptive treatment strategies that dynamically adjust drug dosage based on patients' real-time tumor burden. We employed Particle Swarm Optimization (PSO) to determine the optimal instantaneous dosage for individual patients. The objective function $Q$, defined below, allows the algorithm to minimize the growth of heterogeneity in tumor subpopulations between assessment intervals:
\begin{equation}
Q = \sum_{n=1}^{4} k_i(x_i(t+\bigtriangleup t)-x_i(t)),
\end{equation} 
where $x_i = \left\{ x_1, x_2, x_3, x_4 \right\}$ represents the population size of four distinct tumor cell subclones, $\Delta t$ is the dosing interval (7 or 21 days), and $k_i$ denotes the weighting factors assigned to each subclone (set as $1$, $2$, $2$, $4$ to penalize aggressive phenotypes). 

Consistent with the fixed-schedule analysis, we evaluated the impact of two key variables on adaptive therapy performance: 
\begin{itemize}
\item Dosing interval: The frequency of dose adjustment (7 days vs. 21 days).
\item Maximu permissible dose: The upper limit of the PSO search space, set to $1\times$, $1.5\times$, and $2\times$ the standard clinical dose.
\end{itemize}
For each strategy, we simulated $1,000$ virtual patients to analyze both the recurrence rate and the probability distribution of time to recurrence (TTR).

\begin{figure}[htbp]
\centering
\includegraphics[width=0.9\linewidth]{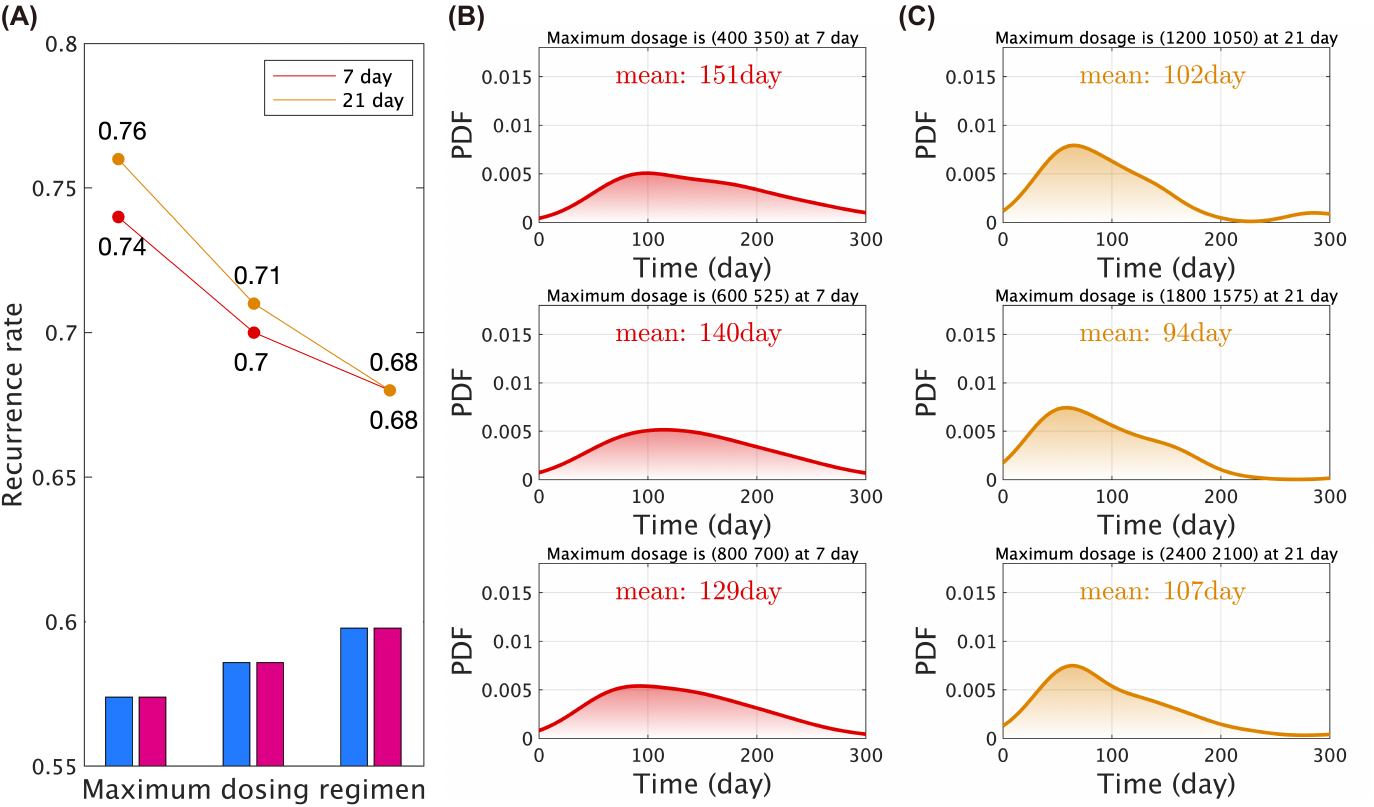}
\caption{{\bf Investigation of recurrence rates and time to recurrence (TTR) under PSO-driven adaptive therapy.} \textbf{(A)} Recurrence rates under different dosing intervals (7 vs. 21 days) and maximum dose limits. The bar charts illustrate results for different maximum search doses (standard, $1.5\times$, $2\times$); blue bars represent atezolizumab and purple bars represent bevacizumab levels. \textbf{(B-C)} Probability distribution function (PDF) of recurrence time for patients who relapse under a 21-day cycle (B) and a 7-day cycle (C). ``Mean'' indicates the average time to recurrence in days.}
\label{fig:9}
\end{figure}

The simulation results revealed a complex trade-off between tumor suppression and evolutionary dynamics. Increasing the maximum permissible dose effectively reduced the overall recurrence rate (the percentage of patients who relapsed) (Fig. \ref{fig:9}A). Shortening the dosing cycle from $21$ to $7$ days resulted in a slight increase in the recurrence rate, indicating a minor drawback of more frequent dosing adjustments. Notably, when the maximum dosage was increased to $2\times$, the impact of the dosing cycle on the recurrence rate was minimized.

However, a different trend emerged when analyzing the time to recurrence (TTR) for those patients who did relapse. We compare the TTR distributions under 21-day (Fig. \ref{fig:9}B) and 7-day (Fig. \ref{fig:9}C) cycles and found that patients in the 7-day cycle group exhibited significantly longer TTR. At the standard maximum dose ($1\times$), the mean TTR was prolonged in the 7-day cycle group ($156$ days) compared to the 21-day group ($109$ days). This indicates that more frequent adaptive adjustments (7-day cycle) can delay disease progression by approximately $45$ days, offering a substantial clinical survival benefit.

Conversely, increasing the maximum permission dose--while reducing the total number of relapsing patients--tended to accelerate relapse in those who failed treatment. As the maximum dose increased, the mean TTR advanced (shortened) by $6$ days in the 21-day group and by 13 days in the 7-day group. This phenomenon suggests that higher doses may exert strong selective pressure, eliminating sensitive cells rapidly but potentially accelerating the competitive release of resistant subclones.

In conclusion, the QCIC model demonstrates high sensitivity to the dosing cycle. A 7-day adaptive cycle with a standard dose limit appears to be the optimal strategy for extending progression-free survival (delayed recurrence), whereas higher doses provide stronger initial suppression but may lead to faster failure upon relapse.  
	
\subsection{Identification of predictive biomarkers for advanced RCC utilizing the QCIC model} 
\label {sec3-6}
	
To identify potential biomarkers predicting disease progression in patients with advanced RCC, we analyzed 9 immune microenvironment indicators in the virtual patient cohort. These indicators included the cell densities of four distinct populations--helper T cells (Th), regulatory T cells (Tr), cytotoxic T cells (Tc), and tumor-associated macrophages (TAM)--as well as five corresponding ratios: Th/Tr, Th/Tc, Th/TAM, Tr/TAM, and Tc/TAM.

We used numerical stimulation data from the second clinical diagnosis point (Week 6) to predict long-term treatment outcomes (one year post-treatment). Binary classification models were employed to construct receiver operating characteristic (ROC) curves for each biomarker.

\begin{figure}[htbp]
\centering
\includegraphics[width=0.9\linewidth]{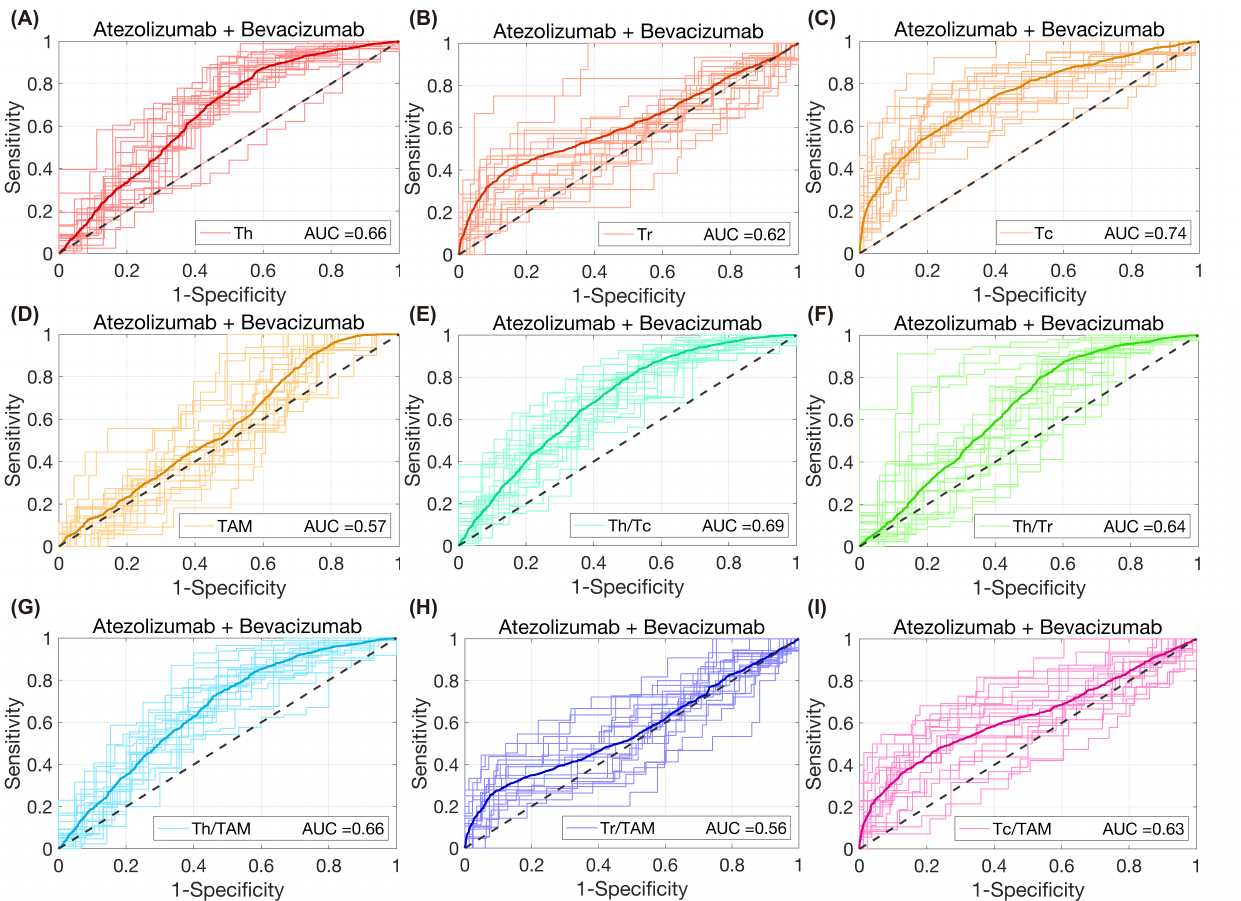}
\caption{{\bf ROC analysis of predictive biomarkers for combination therapy outcomes.} The receiver operating characteristic (ROC) curves illustrate the performance of individual biomarkers in predicting one-year treatment response (Responder vs. Non-responder) based on Week 6 simulation data. \textbf{(A-D)} ROC curves for individual cell densities: Th (A), Tr (B), Tc (C), and TAM (D). \textbf{(E-I)} ROC curves for cell ratios: Th/Tr (E), Th/Tc (F), Th/TAM (G), Tr/TAM (H), and Tc/TAM (I). The area under the curve (AUC) values are annotated in each subplot, with Tc density (C) showing the highest diagnostic accuracy.}
\label{fig:10}
\end{figure}

As shown in Fig. \ref{fig:10}, predictive performance varied significantly across indicators. Among the 9 candidates, tumor-infiltrating Tc cell density exhibited the highest predictive accuracy for combination therapy, achieving an area under the curve (AUC) of $0.74$ (Fig. \ref{fig:10}C). This suggests that the density of effector T cells at the early stage of treatment (Week 6) is a critical determinant of long-term therapeutic success. In contrast, biomarkers related to tumor-associated macrophage (TAM) showed poor predictive value. Notably, the Tr/TAM ratio showed the lowest performance with an AUC of $0.56$ (Fig. \ref{fig:10}H), indicating near-random prediction. Similarly, other TAM-related ratios failed to provide significant prognostic information.
	
In conclusion, our model identifies tumor-infiltrating Tc cell density as a robust biomarker for predicting one-year treatment outcomes, whereas TAM density and its associated ratios appear unsuitable as standalone predictors in this specific therapeutic context. 

\section{Discussion}
\label{sec4}
	
With the rapid advancement of precision medicine, the development of combination therapy regimens has become a cornerstone of oncological research. However, the high cost, prolonged duration, and uncertain efficacy of clinical trials, compounded by significant inter-patient heterogeneity, make it challenging for standard protocols to meet individual therapeutic needs. Consequently, optimizing dosing regimens to maximize therapeutic efficacy has emerged as a critical imperative. In this study, we constructed a Quantitative Cancer-Immunity Cycle (QCIC) model based on multi-scale mechanistic principles to generate a virtual patient cohort that mirrors clinical characteristics. Utilizing this platform, we systematically evaluated treatment outcomes under various dosing strategies. Our findings indicate that a regimen combining short dosing cycles, high-dose bevacizumab,  and low-dose atezolizumab effectively reduces recurrence rates. Furthermore, for patients with a history of recurrence, adaptive dosing strategies showed significant potential to delay disease progression.

Building upon previous quantitative systems pharmacology (QSP) frameworks, this study represents a significant methodological advancement. Want et al. previously pioneered the exploration of dynamic disease progression and drug treatment by integrating biological system mechanisms \citep{Wang.JITC.2021,Wang.iScience.2022}. Compared to their work, our QCIC model incorporates a more comprehensive array of immune cell subsets and cytokines, introducing multidimensional factors for a holistic analysis. Additionally, we adopted the inclusion probability method proposed by  \citet{Allen.CPT.2016}, which serves as an effective tool for filtering virtual patients to match statistically observed data (e.g., from the iAtlas portal). However, a limitation of previous implementations was the lack of clearly defined physiological ranges prior to probability calculation. In our study, we addressed this by integrating key patient characteristics---such as tumor growth rate, dendritic cell-antigen binding efficiency, and immune cell differentiation ratios---alongside the four standard immune subsets. While these features exhibit significant heterogeneity and are currently difficult to quantify directly from immunogenomic data, our model successfully reproduces immune heterogeneity and cell-proportion distributions that closely align with clinical observations in renal cell carcinoma. This validation underscores the model's reliability as a tool for exploring mechanism-driven personalized treatment strategies.

Despite these promising results, several limitations inherent to the current model framework must be acknowledged.  First, regarding pharmacokinetic (PK) parameterization, we utilized a standard body weight ($75$ kg) for dose standardization due to the absence of individual weight data in the iAtlas dataset. Second, the model simplifies certain biological complexities; for instance, it currently does not account for the polarization dynamics of tumor-associated macrophages (TAMs) (M1 vs. M2 phenotype) or detailed intracellular molecular regulatory networks. Future iterations of the model will aim to integrate these macrophage polarization dynamics and deepen the systematic characterization of intracellular mechanisms. Third, the scarcity of longitudinal, individual-level clinical data due to confidentiality restrictions meant that model fitting was primarily performed at the population level rather than the individual level.
	
Future research directions will focus on overcoming these data and mechanistic limitations. By combining machine learning algorithms with multi-omics data mining, we aim to predict model parameters with higher precision. Furthermore, collaborating with clinical institutions to obtain de-identified, longitudinal individual patient data would significantly enhance the model's fitting accuracy. Such advancements would facilitate the transition from population-based simulations to true ``digital twins,'' thereby supporting the formulation of highly precise, personalized clinical decision-making systems. 

\backmatter

\bmhead{Acknowledgements}

This work was funded by the National Natural Science Foundation of China (Grant No. 12331018).

\bmhead{Data and Code Availability} The experimental data are available from references. The source code supporting the findings of this study is available on GitHub: \texttt{https://github.com/jinzhilei/QCIC-RCC}.

\bibliography{references}
	
\end{document}